\documentclass[aps, prl, 10pt, superscriptaddress, twocolumn]{revtex4-1}
\usepackage{amsmath,amssymb,mathrsfs}
\usepackage{graphicx} 
\usepackage{psfrag}
\usepackage[caption=false]{subfig}
\usepackage{bm}

\newcommand{\la}{\label}
\newcommand{\be}{\begin{equation}}
\newcommand{\ee}{\end{equation}}
\newcommand{\bea}{\begin{eqnarray}}
\newcommand{\eea}{\end{eqnarray}}
\newcommand{\p}{\partial}
\newcommand{\1}{\frac{1}{2}}

\newcommand{\comment}[1]{}

\usepackage{color}

\newcommand{\half}{\frac{1}{2}}
\allowdisplaybreaks[1]

\begin{document}
\title{ Investigating anisotropic quantum Hall states with bimetric geometry}


\author{Andrey Gromov}
\email{gromovand@uchicago.edu}
\affiliation{Kadanoff Center for Theoretical Physics, University of Chicago, Chicago, Illinois 60637}


\author{Scott D. Geraedts}
\affiliation{Department of Electrical Engineering, Princeton University, Princeton, New Jersey 08544}

\author{Barry Bradlyn}
\email{bbradlyn@princeton.edu}
\affiliation{Princeton Center for Theoretical Science, Princeton University, Princeton, New Jersey 08544}

\begin{abstract}
We construct a low energy effective theory of anisotropic fractional quantum Hall (FQH) states.  We develop a formalism similar to that used in the bimetric approach to massive gravity, and apply it to describe abelian anisotropic FQH states in the presence of external electromagnetic and geometric backgrounds. We derive a relationship between the shift, the Hall viscosity, and a new quantized coupling to anisotropy, which we term \emph{anisospin}.  We verify this relationship by numerically computing the Hall viscosity for a variety of anisotropic quantum Hall states using the density matrix renormalization group (DMRG). Finally, we apply these techniques to the problem of nematic order and clarify certain disagreements that exist in the literature about the meaning of the coefficient of the Berry phase term in the nematic effective action.
\end{abstract}

\maketitle

\paragraph{Introduction.}

 In recent years there have been a plethora of new advancements in the physics of fractional quantum Hall effect. Notably, several related developments  involving the interplay of quantum Hall physics and geometry have emerged. 
 First, it was recently understood that the response of a quantum Hall state to variations of the background spatial geometry reveals universal properties of the state that go beyond topological effective theory\cite{WenZeeShiftPaper, lee1994orbital,  klevtsov2014random, Klevtsov-fields, Abanov-2014, Gromov-galilean, gromov-thermal,GCYFA, can2014geometry, CLW, laskin2015collective, bradlyn2014low, bradlyn2015topological, 2014-ChoYouFradkin, son2013newton, cappelli2016multipole, schine2016synthetic}. A particularly interesting quantity is the Hall viscosity \cite{1995-AvronSeilerZograf, 2009-Read-HallViscosity, 2011-ReadRezayi}, which in rotationally invariant systems is related to the  shift\cite{haldane1983fractional,WenZeeShiftPaper, 2009-Read-HallViscosity}, and is given by a Berry phase accumulated by the quantum Hall wavefunction on a torus under adiabatic changes of the aspect ratio. When rotational invariance is broken, the Hall viscosity becomes a  multicomponent tensor \cite{1998-Avron}, however its properties and relation to Berry phases have not yet been understood. 

 Additionally, there has been a flurry of recent experimental\cite{xia2011evidence, liu2013evidence,yazdaninematic2016, samkharadze2016observation} and theoretical \cite{mulligan2010isotropic, mulligan2011effective, maciejko2013field, you2014theory, regnault2016evidence} interest in quantum Hall states with spontaneously broken rotational symmetry, i.e.~nematic quantum Hall states\cite{balents1996spatially, fradkin1999liquid}.
Because the nematic order parameter is described by a symmetric matrix, it couples to the microscopic degrees of freedom in a way similar to the  background spatial metric \cite{maciejko2013field, you2014theory}. In the isotropic phase, fluctuations of the nematic order parameter are massive and describe the dynamics of the angular momentum $2$ gapped Girvin-Macdonald-Platzman (GMP)\cite{1986-GirvinMacDonaldPlatzman} magnetoroton mode \cite{maciejko2013field}. In the symmetry broken phase the fluctuations of the order parameter are gapless (up to lattice and boundary effects). 

 In this Letter, we develop a unifying formalism that will bridge the chasm between these new areas of quantum Hall physics. We will explain how to construct a low energy effective theory of quantum Hall states with quadrupolar anisotropy, coupled to perturbations of both the electromagnetic field and spatial geometry. Our construction is reminiscent of a bimetric theory of massive gravity\cite{bimetric2010}. The first metric is determined by the geometry of space, while the second metric is determined by the anisotropy.  Note, however, that we treat both metrics as non-dynamical background fields. We will use this formalism to derive (the non-dissipative parts of) linear response coefficients in the presence of anisotropy.  We also introduce a new response function that probes the coupling of a quantum Hall state to anisotropy.  In order to verify our model, we numerically compute the Hall viscosity for a variety of anisotropic quantum Hall states.  Our construction is also well-suited to describe the nematic quantum Hall states in the isotropic phase and with a quenched configuration of the nematic order parameter as discussed in the Supplementary Material\footnote{See Supplementary Material, which contains the additional references \cite{bradlyn-read-2012kubo,LopezFradkin,1989-Read}}. 

\paragraph{Geometry.}

We start with a brief review of the geometry relevant in quantum Hall physics. Spatial geometry is described by a set of vielbeins  -- or frame fields -- $e_\mu^A=\{{e_0^A\equiv 0},e_i^A\}$ along with their ``inverses'' $ E^\mu_A$ \footnote{we restrict $e^0_\mu$ and $E_0^\mu$ to be trivial.}. Here and throughout we use $\mu,\nu=0,1,2$, and $i,j=1,2$ to index ambient spacetime and space respectively, while $A,B=1,2$ will index flat internal space. The spatial metric $g_{ij}$ is given as
\begin{equation}
g_{ij}=e_i^Ae_j^B\delta_{AB}. \label{eq:metricdef}
\end{equation}
Parallel transport in spacetime is defined by demanding that the vielbeins are covariantly constant, i.e.
\begin{align}
  \nabla_\mu e_\nu^A= \p_\mu e_\nu^A - \Gamma^\lambda_{\mu\nu} e^A_\lambda +  \omega^A_{\mu B} e^B_\nu=0\,,\label{eq:covconst}
\end{align}
where $\nabla_\mu$ is a covariant derivative with a \emph{spacetime} index. These equations define both a spin connection $\omega_\mu^A{}_B$ and a Christoffel connection $\Gamma^i_{\mu j}$.  The Christoffel connection can be expressed in terms of derivatives of the metric, although we will not need the explicit expression here. Solving Eq.~\eqref{eq:covconst} for the spin connection, we find \footnote{We set the ``reduced torsion''\cite{bradlyn2014low} to zero.}
\bea
\omega_0&=&  \frac{1}{2}\epsilon_A{}^B\left(E^i_B\partial_0 e_i^A\right)\,,\label{eq:omega0}\\
\omega_k&=&  \frac{1}{2}\epsilon_A{}^B\left( E^i_B\partial_k e_i^A - E^i_Be^A_j\Gamma^j_{k i}\right) \label{eq:omegak}\,,
\eea
where we have defined the abelian spin connection via $\omega^A_\mu{}_B \equiv \epsilon_B{}^A \omega_\mu$, where $\epsilon_B{}^A$ is the Levi-Civita  tensor.

Lastly, we review the transformation laws for these geometric fields. First, we note that the metric, the vielbeins, and the spin connection all transform as tensors under changes in the ambient coordinates (here we restrict to transformations that leave time invariant). Next, since the vielbeins are defined through the factorization Eq.~(\ref{eq:metricdef}), they suffer an $SO(2)$ gauge ambiguity
\begin{equation}
e^A_\mu\rightarrow e^{A^\prime}_\mu S_{A^\prime}{}^A\,,\qquad  E_A^\mu \rightarrow  E_{A^\prime}^\mu S^{A^\prime}{}_A\,,
\end{equation}
for $S=\exp(i\varphi\epsilon)\in SO(2)$; the spin connection transforms under rotations as an abelian gauge field
\be
\omega_\mu \rightarrow \omega_\mu + \p_\mu\varphi\,.\la{eq:omegatrans}
\ee

\paragraph{Anisotropic geometry.}
Anisotropy naturally arises in condensed matter systems through symmetric rank two tensors, such as the effective mass tensor or dielectric tensor in crystals. Taking inspiration from this, we will introduce anisotropy into the quantum Hall system through a symmetric tensor $V$, distinct from the spatial metric tensor $g$. To be consistent, we must be careful to account for the difference between spatial geometry -- which we view as extrinsically imposed -- and anisotropy -- which we view as intrinsic.  Our discussion here elaborates on and {extends} various observations made in Refs.~\cite{maciejko2013field, you2014theory} and is the first result of the Letter.
 
 We will choose a fairly general type of anisotropy parametrized by a quadrupolar background field $V^{AB}(\mathbf{x})$ which we take to have \emph{internal} $SO(2)$ indices. We require that $V^{AB}$ is symmetric and positive-definite.  We also define the inverse matrix $v_{AB}$ satisfying
  \begin{equation}
  v_{AB}V^{BC}=\delta^C_A
  \end{equation}
 Without loss of generality we can fix $\det V=1$; changes to the determinant of $V$ can be compensated by a uniform rescaling of coordinates, which would not introduce any anisotropy. In analogy with the spatial metric, we can factorize $V$ and $v$ as{
\bea \label{eq:Tdef}
V^{AB}= \Lambda^A_\alpha \Lambda^B_\beta \delta^{\alpha\beta}\,,\quad v_{AB}=\lambda_A^\alpha\lambda_B^\beta\delta_{\alpha\beta}
\eea
}
Note that the indices $\alpha,\beta=1,2$ appearing in Eq.~(\ref{eq:Tdef}) are a new type of internal index. Rotations acting on this index are a new gauge redundancy, distinct from the internal $SO(2)$ rotational symmetry of the previous section. In order to distinguish between these two gauge groups, we will refer to the new redundancy in the {description of} anisotropy as $\widehat{SO}(2)$. 

It is natural to 
define an \emph{anisotropy metric} 
\begin{equation}
\hat{g}_{ij}\equiv e^{A}_i e^{B}_jv_{AB}=\delta^{\alpha\beta}e_i^A \lambda_A^\alpha e_j^B\lambda_B^\beta =\delta^{\alpha\beta}\hat e_i^\alpha  \hat e_j^\beta\label{eq:hatmetricdef}\,,
\end{equation}
where we have introduced $e_i^A \lambda_A^\alpha  = \hat e_i^\alpha$.  We similarly define the inverse
\begin{equation}
\hat{G}^{ij}\equiv\hat E^i_A \hat E^j_B\delta^{AB}.\label{eq:hatmetricinvdef}
\end{equation}
 With two metrics around, we must be careful to distinguish between tensor fields and their inverses. We use the convention that \emph{for the spatial metric only} $g^{ij}g_{jk}=\delta^i_k$. Spatial indices are raised and lowered by this metric, while internal indicies $A$ and $\alpha$ are both raised and lowered by $\delta$ symbols. It {would be} a grave error to use $\hat{g}$ or $\hat{G}$ to manipulate indices.

 The anisotropy {data} $\hat{g}$ and $\hat{e}$ can be used to construct connections and curvatures, just like their  geometric relatives from the previous Section. Any description of an anisotropic system in terms of $\hat e$  with fully contracted indices will \emph{automatically} be spatially covariant.  In particular, we may define a hat-covariant derivative $\hat{\nabla}$ satisfying
\begin{equation}
\hat{\nabla}_\mu\hat{g}_{ij}=0\,, \qquad \hat{\nabla}_\mu\hat{e}^{\alpha}_j=0\,.
\end{equation}
This defines for us implicitly an affine connection $\hat{\Gamma}$, as well as an $\widehat{SO}(2)$ spin connection $\hat{\omega}$ given by replacing all factors of the metric and vielbeins in Eqs.~(\ref{eq:omega0})--(\ref{eq:omegak}) with their hatted cousins.
Clearly, $\hat{\omega}$ transforms as an abelian $\widehat{SO}(2)$ gauge field  under rotations in the internal $\{\alpha,\beta\}$ space, in analogy with Eq.~(\ref{eq:omegatrans}).

Given these two geometries, we define a matrix-valued one-form  $C^i_{\mu j}$ 
\begin{equation}
{C^i_{\mu j}= \Gamma^{i}_{\mu j}-\hat{\Gamma}^i_{\mu j}\,.}
\end{equation}
We also define $C_\mu = \epsilon_i{}^j C^i{}_{j\mu}$ for future use. There are no more independent objects. 

 {The cohomology class (or, informally, the Chern number) $\hat \chi = \frac{1}{2\pi} \int d\hat\omega$ is not independent of the Euler characteristic $\chi$. We find}
\be
\hat \chi = \frac{1}{4\pi} \int \sqrt{\hat g} \hat R = \chi + N_{\rm discl}\,,
\ee
 for some integer $N_{\rm discl}$. Indeed, taking $V^{AB} = \delta^{AB}$ we have $\hat \omega_\mu = \omega_\mu $, and  so $\hat{\chi}
 = \chi$. On the other hand, when the metric (can be and) is set to identity $g_{ij} = \delta_{ij}$ we find

\be
N_{\rm discl} = \frac{1}{2\pi} \int d \hat \omega \Big|_{g_{ij} = \delta_{ij}} \,,
\ee
where $N_{\rm discl}$ counts the number of singularities of the anisotropic connection.  When $V^{AB}$ comes from a nematic order parameter, this integer is related to the number of nematic disclination defects.

\paragraph{Anisotropic Chern-Simons theory.}

We now consider a generic abelian, {anisotropic} one-component FQH system {in a curved space, coupled to a weak external electromagnetic field}. The low energy theory for such a phase is a $U(1)_k$ Chern-Simons action coupled to our anisotropy connections, with $k=2p+1$. Note that an anisotropy tensor $V^{AB}$ can be generated dynamically from the interplay between the dielectric tensor, band mass curvature, in-plane magnetic field, quadrupolar interactions, etc. The only assumption we make is that such a $V^{AB}$ exists. Although we will primarily be interested in cases where the spatial metric is flat or nearly flat, we must first formulate the theory in a general background, so as not to miss any allowed couplings{, nor introduce prohibited ones}.

Given our previous discussions, the most general low energy effective action is \footnote{Although one could include couplings to $\mathrm{Tr}(V)$, they do not contribute to the quantities of interest, and hence we disregard them here and throughout.}
\be\la{eq:Seff}
S = \frac{{2p+1}}{4\pi} \int ada - \frac{1}{2\pi} \int \mathcal A da\,,
\ee  
where
\be
\mathcal A_\mu = A_\mu + s \omega_\mu + \varsigma \hat \omega_\mu +\xi C_\mu\,.
\ee
The coefficients $s$ and $\varsigma$ must be quantized because $\omega$ and $\hat \omega$ are connections  (as opposed to one-forms), however $\xi$ can be an arbitrary function of the anisotropy. For small anisotropy we expect $\xi$ to be approximately independent of the anisotropy, and we focus on this situation throughout the remainder of the text.  We note that a nonzero $\xi$ explicitly breaks the apparent symmetry between the ambient and anisotropy metrics, and cannot be excluded on the basis of effective field theory.

Supplementing the action \eqref{eq:Seff} with an appropriate gauge-fixing condition \cite{GCYFA}, we  integrate out $a$ to derive the generating functional \footnote{As an example, in the Supplementary Material, we derive the generating function $W$ for a microscopic model of non-interacting electrons with band-mass anisotropy.} 
\be
W =\frac{\nu}{4\pi} \int A d A+ \frac{\nu s}{2\pi} \int A d \omega +  \frac{\nu \varsigma}{2\pi} \int A d \hat\omega  + \frac{\nu  \xi}{2\pi} \int A d C\,, \label{eq:genfunc}
\ee
where $\nu =1/(2p+1)$ and we have dropped purely gravitational terms.
The electric charge density is given by
\be
\rho = \frac{\nu}{2\pi} B + \frac{\nu s}{4\pi} R + \frac{\nu \varsigma}{4\pi} \hat R + \frac{\nu \xi}{4\pi} \epsilon^{ij} \p_i C_j\,,
\ee
which implies the total particle number on a sphere
\be
\nu^{-1} N =  N_\phi + \mathcal S +  \varsigma N_{\rm discl}\,,\label{eq:shift}
\ee
where $\mathcal S = 2\bar s=2(s+\varsigma)$ is the {\it shift} \cite{haldane1983fractional,WenZeeShiftPaper}. We see that anisotropy provides a natural way to split the mean orbital spin $\bar s$ into two parts: one that comes from the geometric spin and another one that couples to anisotropy. Thus we will refer to $\varsigma$ as \emph{anisospin} by analogy.  To remind the reader that the anisospin bears a resemblance to the ordinary orbital spin, we denote it by $\varsigma$, the Greek ``final Sigma''.

\paragraph{Hall viscosity and response to anisotropy.}
 We next consider the response of the stress tensor to applied strains, defined from the generating functional as 
\begin{equation}
\tau^{\mu}_{\hphantom{\mu}A}=\frac{\delta  W}{\delta e_\mu^A}=\lambda^{\mu\hphantom{A}\lambda}_{\hphantom{\mu}A\hphantom{\lambda}B}e_\lambda^B+\eta^{\mu\hphantom{A}\lambda}_{\hphantom{\mu}A\hphantom{\lambda}B}\partial_0e_\lambda^B\,.
\end{equation}
We focus on the non-dissipative Hall viscosity 
\begin{equation}
{(\eta^H)}^{\mu\hphantom{A}\lambda}_{\hphantom{\mu}A\hphantom{\lambda}B}=
\frac{1}{2}\left(\eta^{\mu\hphantom{A}\lambda}_{\hphantom{\mu}A\hphantom{\lambda}B}-\eta^{\lambda\hphantom{B}\mu}_{\hphantom{\lambda}B\hphantom{\mu}A}\right).
\end{equation}
In the rotationally invariant case, it has one independent component $\eta^H=\mathcal{S}\bar \rho/4$, proportional to the shift\cite{1998-Avron,2009-Read-HallViscosity}, and the average density $\bar{\rho}$. In the presence of anisotropy however, two new -- and in general non-universal -- contributions to the viscosity tensor emerge. To study these, we follow Haldane and introduce the contracted Hall tensor \cite{haldane2015geometry,haldane2009hall}
\begin{equation}
\eta^H_{AB} = {\half}\epsilon^{CD}e^{A^\prime}_\mu e^{B^\prime}_\nu \epsilon_{A^\prime A} \epsilon_{B^\prime B} \eta^\mu{}_C{}^\nu{}_D.
\end{equation}
From the generating functional Eq.~\eqref{eq:genfunc} we find \footnote{We have assumed here that the density of disclinations vanishes in the thermodynamic limit, and that the characteristic length scale for variations of the anisotropy is large compared to the magnetic length.}

\be
{\eta^H_{AB}= \frac{\bar\rho}{2}\left[s\delta_{AB} + \varsigma  v_{AB} +{\xi} \left(v_{AC}v_{CB} - \delta_{AB}\right)\right]} \,.\label{eq:halltensor}
\ee
In the isotropic limit this reduces to $\eta^H_{AB} = \eta^H\delta_{AB}$. We see that the Hall viscosity and the shift are only proportional in the special cases that either the anisotropy, or both $\varsigma$ and $\xi$ vanish. The contributions from $\xi$ and $\varsigma$ can be distinguished through their scaling with $V$.  In the Supplementary Material, we use the Kubo formalism to derive the Hall tensor $\eta^H$ for a microscopic model of non-interacting electrons with band-mass anisotropy.

{The anisospin $\varsigma$ can also be calculated independent of $s$ via the response to anisotropy. To see this, we define an ``anisotropy  current'' \cite{you2014theory}
\be
\mathscr N^A_\alpha =  \frac{1}{2\lambda}\frac{\delta W}{\delta \lambda_A^\alpha}\,,
\ee
where $\lambda = {\mbox \rm det} (\lambda^\alpha_A)$. Following the logic of Eqs.~(\ref{eq:genfunc}--\ref{eq:halltensor}), we find for the (contracted) odd part of the response of $\mathcal{N}$ to $\partial_0\lambda^\alpha_A$,
\be
{\vartheta^H_{AB} = \frac{\bar\rho}{2}\left(\varsigma + {\xi} \right) v_{AB}}\,.
\ee
Note that $\vartheta^H_{\alpha \beta}$ contains only $\varsigma$ and $\xi$, but not $s$. }

\paragraph{Anisospin for realistic systems.}

Next, let us consider the case where anisotropy enters through the band mass tensor $m^{-1}_{ij}$, and through a distortion of the interaction potential.
\be\la{eq:Ham}
H = m^{-1}_{ij} \pi_i \pi_j + U\left(|x-x^\prime|; \varepsilon_{ij}\right)\,,
\ee
where $\pi_i$ is the momentum (independent of the anisotropy) and $U(|x-x^\prime|;\varepsilon_{ij})$ is the Coulomb potential in a medium with a homogeneous  -- but not necessarily isotropic -- dielectric tensor $\varepsilon_{ij}$\cite{landau1961electrodynamics,Yang2017}. We will assume that these tensors are diagonal and unimodular, with $m^{-1}_{ij} = {\rm diag}(\alpha_m, \frac{1}{\alpha_m})$ and $\varepsilon_{ij} = {\rm diag}(\alpha_\varepsilon, \frac{1}{\alpha_\varepsilon})$. 

To simplify the problem we can make a global coordinate rescaling to move all of the anisotropy into the interaction.  We are then left in Eq.~(\ref{eq:halltensor}) with a single matrix $ v_{ij} = \varepsilon_{ik} m_{kj} = {\rm diag}(\frac{\alpha_\varepsilon}{\alpha_m}, \frac{\alpha_m}{\alpha_\varepsilon})$. 
 Next, note that each cyclotron orbit in the $N$-th Landau level carries orbital angular momentum 
 \be
 s_N=(2N-1)/2
 \la{eq:sn}
 \ee
about its guiding center, an effect which originates from the now-isotropic kinetic term. Hence, for FQH states in the $N$-th Landau level we expect for the geometric spin $s=s_N$. This implies 
\be
\varsigma=\mathcal{S}_N/2-s_N,
\la{eq:sigman}
\ee
where $\mathcal S_N$ is the shift for the state in the $N$-th Landau level. We see that the shift $\mathcal S$ decomposes into cyclotron and interaction contributions.

  Alternatively, we could have rescaled the interaction to move the anisotropy into the band mass tensor. This would lead to a different matrix $ v'_{ij} =  m^{-1}_{ik}\varepsilon^{-1}_{kj} = {\rm diag}(\frac{\alpha_m}{\alpha_\varepsilon}, \frac{\alpha_\varepsilon}{\alpha_m})$. However, note that the anisotropy is a coordinate on the real projective line $\mathbb R P^1\approx S^1$, since the overall scale of the Hamiltonian is unimportant.
  The two rescalings $ v$ and $ v'$ correspond to two different coordinate patches covering $\mathbb R P^1$. In the first case the coordinate is $\alpha = \frac{\alpha_m}{\alpha_\varepsilon}$, while in the second case it is $\alpha^\prime = 1/\alpha$. Either coordinate choice is valid away from the poles of $\mathbb RP^1$, and so both parametrizations will produce equivalent results for all observable quantities. The effect of moving all of the anisotropy into the mass tensor will result in swapping the values of $s$ and $\varsigma$, which is consistent with the transformation law of the Hall tensor Eq.~(\ref{eq:halltensor}) under coordinate rescalings.

{\paragraph{Anisotropic momentum polarization.}

We have  numerically calculated the Hall viscosity for a variety of anisotropic quantum Hall states produced from \eqref{eq:Ham}, using DMRG on an infinite cylinder. In this geometry, the Hall viscosity is related to the `momentum polarization' $P_{\rm pol}$, which is the additional momentum in the azimuthal ($x_2$) direction when the cylinder is cut in the $x_1$ direction
\cite{Tu-PhysRevB.88.195412,Zaletel-PhysRevLett.110.236801,Mong2017}. For anisotropic systems,
\begin{equation}
P_{\rm pol}=-\frac{\eta^H_{22}}{2\pi} L^2+O(1).\label{eq:MomentumPolarization}
\end{equation}
The coefficient  $\eta^H_{22}$ is given by Eq.~(\ref{eq:halltensor}). For the anisotropic systems considered here, the extensive part of the momentum polarization will depend on both the orbital spin $s$ and the anisospin $\varsigma$. 
The $O(1)$ constant is related to the central charge\cite{Zaletel-PhysRevLett.110.236801}; studying its response to the anisotropy is more computationally demanding than the $L^2$ term and is an interesting direction for future work.

When a quantum Hall problem projected into a single Landau level is written in a second-quantized basis, interaction anisotropy and mass anisotropy have an identical effect on the matrix elements of the Hamiltonian, and therefore lead to identical ground states in the orbital basis. Therefore we can test both types of anisotropy in \eqref{eq:Ham} in a single simulation.
To compute momentum polarization from these states we first compute the real space entanglement spectrum (RSES) across the cut, and then
average the momentum eigenvalues of all the levels in the RSES, weighted by their entanglement eigenvalues.
 The RSES depends on the shapes of the single-particle orbitals (which are modified by mass anisotropy but not by interaction anisotropy), so the two types of anisotropies will give different results even though the orbital basis wavefunctions are identical.
Additionally, for interaction anisotropy we can compute momentum polarization from Eq.~(8b) of Ref.~\cite{Zaletel-PhysRevLett.110.236801}, which for isotropic single-particle orbitals gives equivalent results for less computational effort.

\begin{figure}[t]
\includegraphics[width=0.5\textwidth]{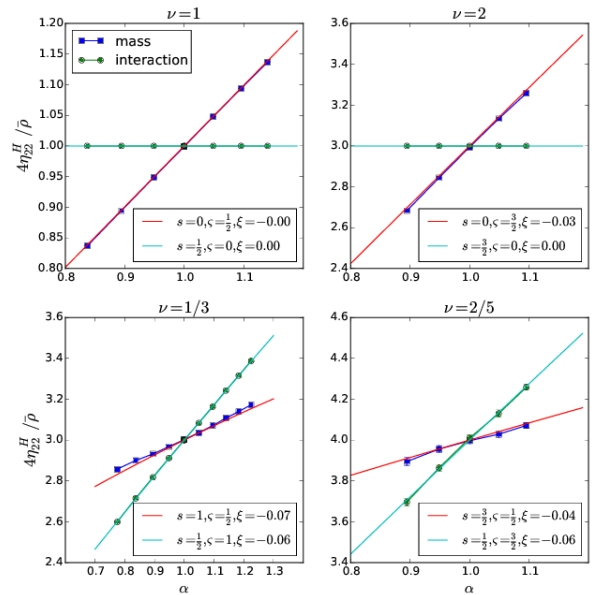}
\caption{Hall viscosity $\eta^H_{22}$ as a function of anisotropy $\alpha$, for four different quantum Hall states. Data is obtained by introducing anisotropy into either the mass (blue squares) or interaction (green circles) part of the Hamiltonian. The lines correspond to Eq.~(\ref{eq:halltensor}),using the values of $s$ and $\varsigma$ given in Eqs.~(\ref{eq:sn}-\ref{eq:sigman}), but allowing $\xi$ to fluctuate to fit the data. The value at $\alpha=1$ is the shift ($\mathcal S_N$). Data obtained using system sizes $L=10-20$ and bond dimensions up to $5400$. The data is plotted such that in the isotropic case it is equal to the shift.}
\label{fig:modS}
\end{figure}

We compute the Hall viscosity by fitting the computed momentum polarization vs. circumference $L$ for a number of different system sizes. In Fig.~\ref{fig:modS} we show results for the integer quantum Hall effect with $\nu=1,2$, the Laughlin state with $\nu=1/3$, and the hierarchy state at $\nu=2/5$. 
The solid curves are fits to Eq.~(\ref{eq:halltensor}), where $s$ and $\varsigma$ given by Eqs.~(\ref{eq:sn}-\ref{eq:sigman}),} and $\xi$ is allowed to vary. We find in all cases that the best fit occurs for $-0.1\le\xi\le0$. 
Finite-circumference effects introduce anisotropy-dependent oscillations in $P_{\rm pol}$. The values of $\xi$ we extract may therefore be overestimates, since they could reflect these systematic errors. 
Reducing these finite-size effects would require larger bond dimensions and would (with our present computational resources) simply replace finite-size error with finite-bond-dimension error (finite-bond-dimension error is very small in the data we present).  A finite size scaling analysis (presented in the Supplementary Material), suggests that $\xi$ is small but nonzero.
We have also assumed that $\xi$ is independent of $\lambda$, though this is not required. This, along with finite size effects, may explain the deviations from the fit we observe at large anisotropy in $\nu=1/3$.

We thus see that the effect of anisotropy which couples \emph{only} to either the kinetic energy, or to the interaction potential, is to split the contributions to the shift $\mathcal{S}$ into a single particle ``Landau orbit'' contribution, and a many-body ``guiding center'' contribution. Such a splitting was first noted by Haldane\cite{haldane2009hall,haldane2011geometrical}. The values of $s$ and $\varsigma$ obtained are precisely those suggested in the previous section\footnote{More complicated forms of anisotropy (like those studied in Ref. \cite{haldane2015geometry}) -- described by higher rank tensors --  can be incorporated through a straightforward generalization of the present formalism.}.


\paragraph{Conclusions.}

 We have introduced a framework for studying the low energy properties of anisotropic quantum Hall states. Using it, we constructed a family of low energy theories for anisotropic abelian FQH states, and studied their linear response.  We have found a new quantized topological number $\varsigma$, dubbed anisposin, related to the non-dissipative linear response to anisotropy.  We have shown that in the presence of homogeneous anisotropy the relation between the shift and the Hall viscosity is modified -- while the former remains quantized for any value of the anisotropy (as long as it preserves the inversion symmetry), the Hall viscosity is quantized only in the isotropic case.

We have numerically investigated the Hall viscosity of a variety of quantum Hall states coupled to both band-mass and interaction anisotropy. We have shown that the anisospin for these systems realizes a splitting of the shift between Landau orbit and interaction contributions, first pointed out in Ref.~\cite{haldane2009hall}.

We believe that our formalism will have many applications, including a detailed investigation of the dynamics of gapped collective excitations in FQH systems, nematic phase transitions, and  ``hidden'' geometric degrees of freedom \cite{haldane2011geometrical}. The correspondence between anisotropy and bimetric geometry allows one to construct anisotropic CFT trial states and study corresponding Berry phases,  which we will discuss in a forthcoming work. Finally, our geometric description may help to build a bridge between FQH physics and bimetric theories of massive gravity.

\begin{acknowledgments}
{\it Acknowledgements.} We are pleased to thank A. Abanov, E. Fradkin, J. Maciejko, Z. Papic, N. Regnault, S. Sondhi, B. Yang, and M. Zaletel for useful discussions.
We also acknowledge M. Zaletel and R. Mong for providing the DMRG libraries used for our simulations.
A.G. was supported by the Leo Kadanoff fellowship and the NSF grant DMS-1206648. B.B. and A.G. gratefully acknowledge support from both the Simons Center for Geometry and Physics, Stony Brook University, as well as the Banff International Research Station where some of the research for this paper was performed. S.D.G. is supported by Department of Energy  BES Grant DE-SC0002140.
\end{acknowledgments}


%


\end{document}


\title{Supplementary Material for Investigating anisotropic quantum Hall states with bi-metric geometry}

\author{Andrey Gromov}
\email{gromovand@uchicago.edu}
\affiliation{Kadanoff Center for Theoretical Physics, University of Chicago, Chicago, Illinois 60637}


\author{Scott D. Geraedts}
\affiliation{Department of Electrical Engineering, Princeton University, Princeton, New Jersey 08544}

\author{Barry Bradlyn}
\email{bbradlyn@princeton.edu}
\affiliation{Princeton Center for Theoretical Science, Princeton University, Princeton, New Jersey 08544}
\date{\today}
\maketitle

\section{Anisotropy in a microscopic model}

We consider a microscopic model  of non-interacting electrons in magnetic field with an anisotropic band mass tensor $(m_*^{-1})^{AB}=\frac{1}{2m}V^{AB}$. We will examine the stress response in the IQH regime in two ways. First, we will show directly from the microscopic Hamiltonian that the Hall viscosity is given by Eq.~(23) of the main text. Second, we will derive by direct integration the generating functional of the model, and show that it agrees with Eq.~(17) of the main text.

\subsection{Viscosity from the Kubo Formula}
The Hamiltonian for noninteracting electrons in a magnetic field with this anisotropic band mass tensor is given in flat space by
\begin{equation}
H=\frac{1}{2m}V^{AB}\delta^i_A\delta^j_B\sum_q\pi^q_i\pi^q_j,\label{eq:vischam}
\end{equation}
where $q$ is a particle index, and $\pi^q_i=p_i^q-A_i(\mathbf{x}^i)$ is the kinetic momentum of particle $q$. We will follow the formalism of Ref.~\cite{bradlyn-read-2012kubo} to compute the viscosity tensor from the flat space Hamiltonian. We assume that the coupling to the background metric is minimal, in the sense that the geometry only enters through the replacement
\begin{equation}
\delta^i_A\rightarrow E^i_A,\; \delta^j_B\rightarrow E^j_B
\end{equation}
in Eq.~(\ref{eq:vischam}). For the sake of simplicity, we will consider only non-compressive metric perturbations. We first identify the strain generators 
\begin{equation}
J^i{}_{j}=-\frac{1}{2}\sum_q\left(\left\{x_q^i,\pi^q_j\right\}-B\epsilon_{jk}x_q^ix_q^k\right),\label{eq:straingen}
\end{equation}
which generate area-preserving deformations of the system.
The continuity equation for momentum density implies that the (traceless part of) the integrated stress tensor is given by
\begin{equation}
T^i{}_{j}=-i\left[H,J^i{}_{j}\right].\label{eq:viscward}
\end{equation}
Note that this relationship is precisely a consequence of our choice of geometric coupling. It would be modified if, for instance, we inserted into the Hamiltonian a direct coupling to the tensor $C_\mu$ from the main text.

Using this Ward identity, we can write the non-compressive contributions to the viscosity tensor (compressive contributions will vanish at zero frequency) as
\begin{equation}
(\eta)^{i}{}_j{}^k{}_\ell=\lim_{\omega\rightarrow 0}\frac{1}{V\omega^+}\left(\langle\left[T^i{}_j,J^k{}_\ell\right]\rangle+\langle\langle T^i{}_{j},T^k{}_{\ell}\rangle\rangle\right),\label{eq:visckubo}
\end{equation}
where we have defined the retarded correlation function
\begin{equation}
\langle\langle A,B\rangle\rangle=\int_0^\infty dt e^{i\omega^+t}\langle\left[A(t),B(0)\right]\rangle,
\end{equation}
and all expectation values are taken with respect to the flat-space Hamiltonian Eq.~(\ref{eq:vischam}). While we can proceed directly to evaluate this expression, it is easier to first perform a canonical transformation. Recalling that $V^{AB}=\Lambda^A{}_\alpha\Lambda^B{}_\alpha$, and $\lambda^\alpha{}_A\Lambda^A{}_\beta=\delta^\alpha_\beta$, we define
\begin{equation}
\tilde{\pi}^q_\alpha=\Lambda^A{}_\alpha\delta^i_A\pi^q_i,\; \tilde{x}_q^\alpha=\lambda^{\alpha}{}_A\delta^A_ix^i_q.\label{eq:trans}
\end{equation}
To lighten the notational load, we will exploit the fact that in flat space, index types ($i,A,\alpha$) can be treated equivalently; we will thus from here on out suppress Kronecker $\delta$ symbols which serve only to change index type, with the understanding that the first, upper index of $\Lambda$ is always of $SO(2)$ type ($A$), and the first, upper index of $\lambda$ is always of $\widehat{SO}(2)$ ($\alpha$) type. The constraint $\det(V)=1$ ensures that this transformation is canonical. Under this transformation, the Hamiltonian can be written
\begin{equation}
H=\frac{1}{2m}\sum_{q\alpha} \tilde{\pi}_\alpha^q\tilde{\pi}_\alpha^q,
\end{equation}
which we recognize as equivalent to the \emph{isotropic} Landau Hamiltonian. Furthermore, making use of the inverse transformation
\begin{equation}
\pi^q_i=\lambda^\alpha{}_i\tilde{\pi}^q_\alpha,\;\; x^i_q=\Lambda^i{}_\alpha\tilde{x}^\alpha
\end{equation}

we see that the strain generators can be expressed as

\begin{equation}
J^{i}{}_j=\Lambda^i{}_k\lambda^\ell{}_j\tilde{J}^{k}{}_\ell,
\end{equation}
where $\tilde{J}$ is given by Eq.~(\ref{eq:straingen}) with $\pi,x$ replaced by $\tilde{\pi},\tilde{x}$.
After employing the Ward identity of  Eq.~(\ref{eq:viscward}), we see that the stress tensor is given in terms of the ``tilde''-basis as
\begin{align}
T^i{}_j&=-i\left[H,J^\alpha{}_\beta\right],\nonumber\\
&=\Lambda^i{}_k\lambda^\ell{}_j\left(-i\left[H,\tilde{J}^k{}_\ell\right]\right),\nonumber\\
&=\Lambda^i{}_k\lambda^\ell{}_j \tilde{T}^k{}_\ell,
\end{align} 
where we recognize
\begin{equation}
\tilde{T}^i{}_{j}=\delta^{ik}\frac{1}{2m}\sum_q\left\{\tilde{\pi}^q_k,\tilde{\pi}^q_j\right\}
\end{equation}
as the operator expression for the \emph{isotropic stress tensor}, in terms of $\tilde{\pi}$. With this observation, we have mapped the problem of evaluating Eq.~(\ref{eq:visckubo}) back to the problem of evaluating the viscosity for an isotropic system. Using the known result for the viscosity of an integer quantum Hall system, we find immediately that
\begin{align}
(\eta)^{i}{}_j{}^k{}_\ell&=-\frac{1}{4}\mathcal{S}\rho\left(\delta^{\alpha}_\zeta\epsilon^\gamma{}_\beta-\delta^\gamma_\beta\epsilon^\alpha{}_\zeta\right)\Lambda^i{}_\alpha\lambda^\beta{}_j\Lambda^k{}_\gamma\lambda^\zeta{}_\ell,\\
&=-\frac{1}{4}\mathcal{S}\rho\left(\delta^i{}_\ell\Lambda^k{}_\gamma\epsilon^\gamma{}_\beta\lambda^\beta{}_j-\delta_j{}^k\Lambda^i{}_\alpha\epsilon^\alpha{}_\zeta\lambda^\zeta{}_\ell\right)\label{eq:anisovisc}
\end{align}
where $\mathcal{S}=\nu$ is the shift for $\nu$ filled Landau levels. Noting that
\begin{equation}
\Lambda^k{}_\gamma\Lambda^\ell{}_\beta\epsilon^{\gamma\beta}=\mathrm{det}(\Lambda)\epsilon^{k\ell}=\epsilon^{k\ell}
\end{equation}
implies
\begin{equation}
\Lambda^k{}_\gamma\epsilon^\gamma{}_\beta=\epsilon^{km}\lambda^\gamma{}_m\delta_{\gamma\beta}
\end{equation}
we can simplify Eq.~(\ref{eq:anisovisc}) to find
\begin{equation}
(\eta)^{i}{}_j{}^k{}_\ell=-\frac{1}{4}\mathcal{S}\rho\left(\delta^i{}_\ell\epsilon^{km} v_{jm}-\delta_j{}^k\epsilon^{im} v_{\ell m}\right),
\end{equation}
where $v_{\ell m}=\lambda^\alpha{}_\ell\lambda^\alpha{}_m$.

 Applying formula Eq.~(22) from the main text for the contracted Hall tensor, we thus find
\begin{align}
\eta^H_{AB}&=-\frac{1}{8}\mathcal{S}\rho\epsilon^{j\ell}\epsilon_{iA}\epsilon_{kB}\left(\delta^i{}_\ell\epsilon^{km} v_{jm}-\delta_j{}^k\epsilon^{im} v_{\ell m}\right) \nonumber\\
&=-\frac{1}{8}\mathcal{S}\rho\left(\epsilon^{ji}\epsilon_{iA}\epsilon_{kB}\epsilon^{km}v_{jm}-\epsilon^{k\ell}\epsilon_{iA}\epsilon_{kB}\epsilon^{im}v_{\ell m}\right)\nonumber\\
&=\rho\frac{\mathcal{S}}{4}v_{AB},
\end{align}
and so we deduce that the isospin $\varsigma=\frac{\mathcal{S}}{2}$ for noninteracting electrons. Because of our choice of coupling to the metric, we also find $\xi=0$.

Finally, note that although we considered only non-compressive perturbations for clarity, The canonical transformation Eq.~(\ref{eq:trans}) implies that the \emph{full, frequency dependent} viscosity tensor $\eta(\omega)^{i}{}_j{}^k{}_\ell$ for the anisotropic electron system (including the compressive contributions) is given by
\begin{equation}
\eta(\omega)^{i}{}_j{}^k{}_\ell=\eta_0(\omega)^{\alpha}{}_\beta{}^\gamma{}_\zeta\Lambda^i{}_\alpha\lambda^\beta{}_j\Lambda^k{}_\gamma\lambda^\zeta{}_\ell,
\end{equation}
where, using Ref.~\onlinecite{bradlyn-read-2012kubo}, we have
\begin{align}
\eta_0(\omega)^{i}{}_j{}^k{}_\ell&=\frac{\rho\mathcal{S}\omega_c}{2({\omega^+}^2-4\omega_c^2)}\left[i\omega^+(\delta^{i}{}_\ell\delta^k{}_j-\epsilon^i{}_\ell\epsilon_j{}^k)+2\omega_c\left(\delta^i{}_\ell\epsilon^k{}_j-\delta_j{}^k\epsilon^i{}_\ell\right)\right]
\end{align}
for the frequency-dependent viscosity of the isotropic system.

\subsection{Generating Functional}

In this Section we will directly integrate out the electronic degrees of freedom in the model of non-interacting electrons with anisotropic band mass, in a magnetic field filling $N$ Landau levels.
In flat space the model is described by the action
\be
S =  \int dt d^2x \,\, \left[ i \psi^\dag  D_0 \psi +\frac{1}{2m} \hat V^{AB}  D_A\psi^\dag  D_B\psi\right]\,.
\ee
In curved space one has to replace the covariant derivatives via $D_A = E_A^i D_i$ (see, for example, \cite{bradlyn2014low, gromov2015thermal}), where $E_A^i$ is the inverse spatial vielbein. Then
\be
\frac{1}{2m}  V^{AB}  D_A\psi^\dag  D_B\psi \qquad \longrightarrow \qquad \frac{1}{2m} \hat G^{ij} D_i\psi^\dag  D_j\psi\,.
\ee
The full action in curved space is given by
\be
S=  \int dt d^2x \,\,\sqrt{\hat{g}} \left[ i \psi^\dag  D_0 \Psi +\frac{1}{2m} \hat G^{ij} D_i\psi^\dag  D_j\psi\right]\,.
\ee
We have used $\det V = 1$ to replace $\sqrt{g}$ by $\sqrt{\hat g}$ and $D_\mu =  \p_\mu +iA_\mu$.
 
To integrate out $\psi$ we note that the coupling of $\psi$ to $\hat g$ is identical to the coupling of an isotropic system to the ambient metric. Thus, we can use the results of Ref.~\cite{Abanov-2014} for the effective action. We find
\be
W = \frac{N}{4\pi} \int AdA +  \frac{N^2}{4\pi} \int Ad\hat \omega + \ldots\,,
\ee
which implies $\varsigma = N$ and every Landau level contributes $\varsigma_N = N - \frac{1}{2}$.

Coupling to the tensor $C_\mu$ can be included via the modified covariant derivative.
\be
  D_\mu \rightarrow \p_\mu +iA_\mu+ i\alpha C_\mu\,.
\ee
Since this coupling is the same order in derivatives as the other coupling, we cannot discard it in principle. This coupling leads to the appearence of $\xi = \frac{N}{2} \alpha$ in Eq. (17) of the main text. 

When  inter-election interactions $S_{\rm int}$ are introduced, the couplings to $g$ and $\hat g$ are no longer identical. This leads to the splitting of the shift into $s$ and $\varsigma$ discussed in the main text.

\section{Additional Numerical Data}

In this section we present some additional details about the data in Fig. 1 of the main text. The data in that figure is chosen so that the Hall viscosity has a{constant piece which is equal to $2s_N$, and a piece linear in $\alpha$ with slope $2\varsigma$.} There is also an additional term which goes as $\alpha^2-1$ with coefficient $2\xi$. All this can be seen from Eq.(23) of the main text. In principle, $\xi$ could also depend on $\alpha$, here we are assuming that its dependence on $\alpha$ is constant in the range of $\alpha$ studied. We can use Eq.(26)-(27) to predict the values of $\varsigma$ and $s_N$, how we do this is summarized in Table \ref{table}. Note that these values are the ones used when the anisotropy is in the interaction, when the anisotropy is in the kinetic part of the Hamiltonian $s_N$ and $\varsigma$ are interchanged, as described in the main text. 

\begin{table}
\begin{tabular}{|c|c|c|c|c|}
\hline
$\nu$ & $N$ & $S_N$ & $s_N$ & $\varsigma$ \\
\hline
$1$ & $1$ & $1$ & $\frac12$ & $0$ \\
$2$ & $2$ & $3$ & $\frac32$ & $0$ \\
$\frac13$ & $1$ & $3$ & $\frac12$ & $1$ \\
$\frac25$ & $1$ & $4$ & $\frac12$ & $\frac32$ \\
\hline
\end{tabular}
\caption{This table, combined with Eqs.(26)-(27) in the main text, shows how we can determine the values of $s_N$ and $\varsigma$ used in Fig. 1}
\label{table}
\end{table}

\begin{figure}
\includegraphics[width=0.5\linewidth]{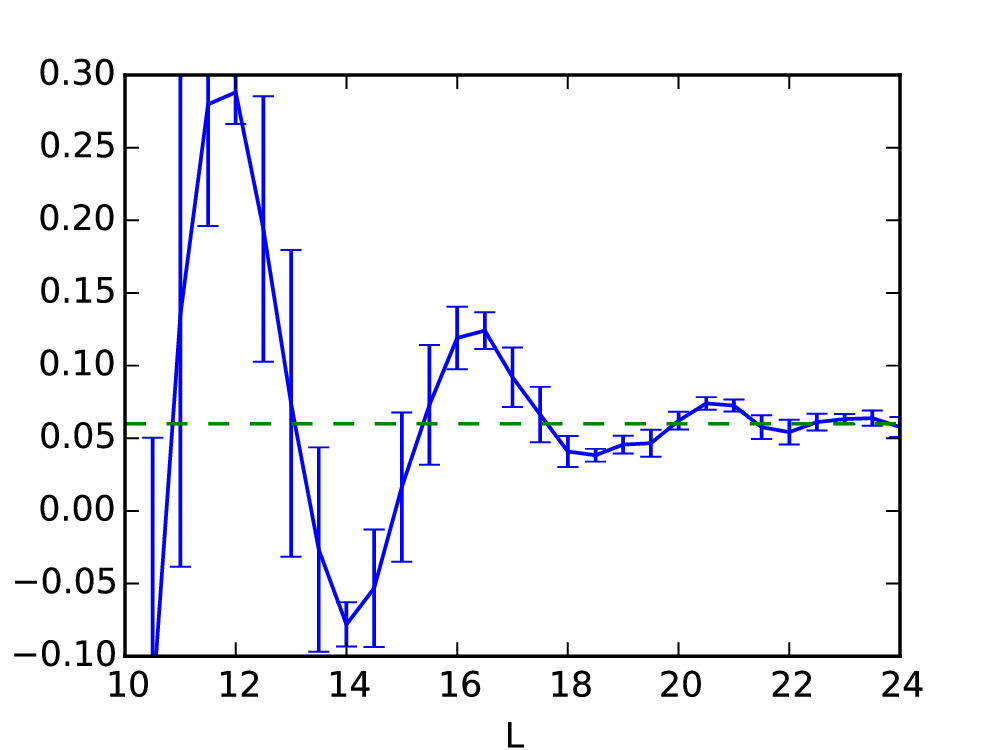}
\caption{Finite size dependence of the value of $\xi$ extracted for the $\nu=1/3$ interaction data in Fig. 1. We see that though there are oscillations as a function of system size, at the largest sizes we can access these oscillations are smaller than the value of $\xi$.}
\label{finitesize}
\end{figure}

In Fig. 1 we find deviations from the straight lines determined by $s_N$ and $\varsigma$, which we attribute to non-zero $\xi$. We might ask if these deviations are, in fact, not from a non-zero $\xi$, but instead arise from the finite circumference of the cylinder. (As explained in the main text,  by going to bond dimensions up to $\chi=8000$ we can reduce finite-bond-dimension errors to $\approx 10^{-3}$, much smaller than the deviations from a straight line seen in the figure). To obtain Fig. 1 for each value of $\alpha$ we fit $\eta_H$ vs $L^2$ to a straight line, using the system sizes $L=10-20$. After doing this for each value of $\alpha$ we obtain the data in Fig. 1 which we fit to obtain $\xi$.  In Fig.~\ref{finitesize} we investigate the finite-size dependence of our $\xi$ values at $\nu=1/3$  by performing the same procedure, but instead of using data at all the sizes we obtain we only use three data points at $L-0.5, L, L+0.5$, for a range of $L$. We see that the data does oscillate as a function of $L$. However, at the largest $L$ we can access it seems that the oscillations on the value of $\xi$ we obtain are smaller than its absolute value, leading us to believe that the non-zero value of $\xi$ we obtain is not a finite-size effect.

\section{Anisotropy from Nematic Order}
As we explained in the main text, the tensor $V^{AB}$ can be generated dynamically. In this Section we will examine a particular way of generating this tensor from the quenched configuration of a nematic order parameter in the isotropic phase. In particular, we will explain how our formalism arises in the context of Ref.~\cite{maciejko2013field} and  Ref.~\cite{you2014theory}, and elucidate an implicit assumption that led to an apparent disagreement. 

In what follows we will {utilize} the computation of Ref.~\cite{you2014theory}. Our starting point is  a system of interacting, non-relativistic electrons with the interaction given by a repulsive Coulomb potential and an attractive quadrupolar interaction:
\be\la{eq:nematicsetup}
S= S_0 +  S_{\rm c} + S_{\rm q}\,,
\ee  
where 
\bea\la{eq:FF}
S_0 &=& \int dt d^2x \,\, \left[ i \psi^\dag D_0 \psi +\frac{1}{2m}\delta^{AB}D_A\psi^\dag D_B\psi\right]\,,\\
S_{\rm c}&=& \int dt \int d^2\mathbf{x} d^2\mathbf{x^\prime} \rho(\mathbf{x}) U(|\mathbf{x}-\mathbf{x^\prime}|) \rho(\mathbf{x^\prime})\,,\\
S_{\rm q}  &=& \int dt \int d^2\mathbf{x} d^2\mathbf{x^\prime} F_2(|\mathbf{x}-\mathbf{x^\prime}|) \tr\Big[Q(\mathbf{x}) Q(\mathbf{x^\prime})\Big]\,,
\eea
where $\rho(\mathbf{x})=\psi^\dag(\mathbf{x}) \psi(\mathbf{x})$ is the electron density and $Q(\mathbf{x})$ is a quadrupolar operator
\be
Q(\mathbf{x}) = \psi^\dag \begin{pmatrix} D_x^2 - D_y^2 & D_xD_y + D_y D_x \\ D_xD_y + D_y D_x   & -D_x^2 + D_y^2 \end{pmatrix} \psi\,,
\ee
 and $D_i = \p_i + i A_i$.

Our goal is to derive the effective action coming from \eqref{eq:nematicsetup} and from it compute the Hall viscosity. To do so, we minimally couple \eqref{eq:nematicsetup} to the spatial metric $g_{ij}$ by replacing $\delta^{AB} \rightarrow \delta^{AB} e_A^i e_B^j$,  introducing a factor of $\sqrt{g}$ in the spatial integration measure, and replacing $|\mathbf{x} - \mathbf{x^\prime}|$ with $d(\mathbf{x} ,\mathbf{x^\prime})$ - the geodesic distance between $\mathbf{x}$ and $\mathbf{x^\prime}$ evaluated with respect to the metric $g_{ij}$. 

Next, we perform the standard flux attachment \cite{LopezFradkin,1989-Read,2014-ChoYouFradkin} by introducing a statistical gauge field $a_i$. We also introduce a Hubbard-Stratonovich field $\mathbf{M} = (M_1,M_2)$ to decouple the quadrupolar interaction $S_{\rm q}$. Then, after doing a mean-field approximation for the attached flux, we find the following effective action:
\be
S_{\rm eff}[\mathbf{M}, \Psi,a] = S_{\rm n}[\mathbf{M}] + \tilde S_0[\Psi, \mathbf{M},a] + \frac{1}{2p} \frac{1}{4\pi} \int ada\,,
\ee
where we have  introduced the composite fermion field $\Psi$, and suppressed external background fields  as arguments.

 The first term  $S_{\rm n}[\mathbf{M}]$ describes the gapped dynamics of the nematic order parameter. We will not need an exact form of this term, it is sufficient to know that it is gapped. The scale at which these massive degrees of freedom become important is the gap of the zero momentum GMP mode, which is of the order of the Coulomb gap. The latter is assumed to be infinite in our low energy theory. Thus we will drop this term for the remainder of the Letter.

 The second term $\tilde S_0[\Psi, \mathbf{M},a]$ describes the dynamics of free composite fermions which we will choose to fill the lowest Landau level
\be\la{eq:CFani}
 \tilde S_0[\Psi, \mathbf{M}]=  \int dt d^2x \,\,\sqrt{\hat{g}} \left[ i \Psi^\dag \tilde D_0 \Psi +\frac{1}{2m} \hat G^{ij}\tilde D_i\Psi^\dag \tilde D_j\Psi\right]\,,
\ee
 where  $\hat G^{ij}$ is given by Eq.~(10) of the main text with $V^{AB}$ given in terms of $\mathbf{M}$ as follows
 \be
 V =  \frac{1}{\sqrt{1-|\mathbf{M}|^2}}\begin{pmatrix} 1+ M_2 & M_1 \\ M_1 & 1-M_2\end{pmatrix}\,,
 \ee
 and $\tilde D_i$ is the covariant derivative that describes the interaction of the composite fermion spin with geometry and anisotropy, which we now discuss. Indeed,  Eq. \eqref{eq:CFani} requires extra explanation and is, in part, the origin of the disagreement between Ref.~\cite{maciejko2013field} and  Ref.~\cite{you2014theory}. The introduction of  a new covariant derivative $\tilde D_i$ is necessary to ensure that flux attachment is an exact transformation even in curved space \cite{lee1994orbital, 2014-ChoYouFradkin}. However, there exists more than one choice of $\tilde D_i$ that will accomplish this task. In fact, the most general $\tilde D_i$ would distribute the coupling between geometry and anisotropy according to $\tilde D_i = D_i +ia_i + ip_1\omega_i + ip_2\hat \omega_i$, with $p_1 + p_2=p$ and $2p_1,2p_2 \in \mathbb Z$. The values of $p_1$ and $p_2$ must be in general determined from a microscopic analysis, which is beyond the scope of the present paper. Thus we will keep the coupling general and use the general  $\tilde D_i$ in the remainder of this Section. 

Integrating out the composite fermions and the statistical gauge field $a$ we find
\be
S_{\rm eff} = \frac{\nu}{4\pi} \int A d \left(p_1 \omega + \frac{2p_2+ 1}{2}\hat \omega\right) +\frac{\nu}{4\pi} \int AdA\,,
\ee
we have suppressed purely gravitational terms. The extra $\frac{1}{2}\hat \omega$ comes from integrating out the composite fermions, which unambiguously couple to $\hat g$. 

To analyze this result we first turn off all external fields, leaving

\bea
S_{\rm eff} &=&  \frac{\varsigma\bar \rho}{2} \int \mathrm  \epsilon_\alpha{}^\beta  \Lambda_\beta^B \p_0 \lambda^\alpha_B + \ldots\,,
\eea
where $\varsigma = \frac{2p_2 +1}{2}$. This term has been obtained in Ref. \cite{maciejko2013field} with anisospin $\varsigma = \bar s = \frac{2p+1}{2}$ and in Ref.~\cite{you2014theory} with $\varsigma =\frac{1}{2}$. These are two opposite cases when $p_2$ is set to be either $p$ or $0$ correspondingly. We see that both scenarios are plausible depending on the value of $p_2$, {\it i.e.} depending on how exactly the spin of a composite fermion couples to anisotropy. We are not aware of a physical criterion that would select a particular value of $p_2$ in the situation when the nematic order parameter is pinned to a fixed configuration. Most likely, this ambiguity points to the fact that the coefficient of the Berry phase term is not universal.

Turning the metric back on we find that Hall viscosity tensor is
\be
\eta^H_{AB} =  \frac{ p_1 \bar \rho}{8} \delta_{AB} + \lambda \frac{ \varsigma \bar \rho}{8} V_{AB} \,,
\ee
where $\lambda = {\mbox \rm det}( \lambda^\alpha_A)$ which reduces to $\eta^H_{AB} = \frac{2p+1}{4} \bar \rho\delta_{AB}$ in the isotropic limit, as it should be. Finally, we note that $\xi=0$ in the flux attachment computation. {It is possible to accommodate a non-zero $\xi$ by coupling the original fermions \eqref{eq:FF} to $C$, however such coupling appears to be unnatural.}

\bibliography{Bibliography-supp}